\documentclass[sigconf,nonacm]{acmart}

\usepackage{algorithm, algpseudocode}
\usepackage{amsmath}
\DeclareMathOperator*{\argmax}{argmax}
\usepackage[autolanguage,np]{numprint}
\usepackage[inline]{enumitem}
\usepackage{microtype}
\usepackage{booktabs}
\usepackage{multirow}

\AtBeginDocument{%
  \providecommand\BibTeX{{%
    \normalfont B\kern-0.5em{\scshape i\kern-0.25em b}\kern-0.8em\TeX}}}

\begin{document}

\title{Building a Scalable, Effective, and Steerable Search and Ranking Platform}


\author{Marjan Celikik}
\email{marjan.celikik@zalando.de}
\affiliation{%
  \institution{Zalando SE}
  \city{Berlin}
  \country{Germany}
}
\author{Jacek Wasilewski}
\email{jacek.wasilewski@zalando.de}
\affiliation{%
  \institution{Zalando SE}
  \city{Berlin}
  \country{Germany}
}

\author{Ana Peleteiro Ramallo}
\email{ana.peleteiro.ramallo@zalando.de}
\affiliation{%
  \institution{Zalando SE}
  \city{Berlin}
  \country{Germany}
}

\author{Alexey Kurennoy}
\email{alexey.kurennoy@zalando.ie}
\affiliation{%
  \institution{Zalando SE}
  \city{Berlin}
  \country{Germany}
}

\author{Evgeny Labzin}
\email{evgeny.labzin@zalando.de}
\affiliation{%
  \institution{Zalando SE}
  \city{Berlin}
  \country{Germany}
}

\author{Danilo Ascione}
\email{danilo.ascione@zalando.de}
\affiliation{%
  \institution{Zalando SE}
  \city{Berlin}
  \country{Germany}
}

\author{Tural Gurbanov}
\email{tural.gurbanov@zalando.de}
\affiliation{%
  \institution{Zalando SE}
  \city{Berlin}
  \country{Germany}
}

\author{Géraud Le Falher}
\email{geraud.le.falher@zalando.de}
\affiliation{%
  \institution{Zalando SE}
  \city{Berlin}
  \country{Germany}
}

\author{Andrii Dzhoha}
\email{andrew.dzhoha@zalando.de}
\affiliation{%
  \institution{Zalando SE}
  \city{Berlin}
  \country{Germany}
}

\author{Ian Harris}
\email{ian.harris@zalando.ie}
\affiliation{%
  \institution{Zalando SE}
  \city{Berlin}
  \country{Germany}
}

\renewcommand{\shortauthors}{Celikik, et al.}

\begin{abstract}

Modern e-commerce platforms offer vast product selections, making it difficult for customers to find items that they like and that are relevant to their current session intent. This is why it is key for e-commerce platforms to have near real-time scalable and adaptable personalized ranking and search systems. While numerous methods exist in the scientific literature for building such systems, many are unsuitable for large-scale industrial use due to complexity and performance limitations. Consequently, industrial ranking systems often resort to computationally efficient yet simplistic retrieval or candidate generation approaches, which overlook near real-time and heterogeneous customer signals, which results in a less personalized and relevant experience. Moreover, related customer experiences are served by completely different systems, which increases complexity, maintenance, and inconsistent experiences.

In this paper, we present a personalized, adaptable near real-time ranking platform that is reusable across various use cases, such as browsing and search, and that is able to cater to millions of items and customers under heavy load (thousands of requests per second). We employ transformer-based models through different ranking layers which can learn complex behavior patterns directly from customer action sequences while being able to incorporate temporal (e.g. in-session) and contextual information. We validate our system through a series of comprehensive offline and online real-world experiments at a large online e-commerce platform, and we demonstrate its superiority when compared to existing systems, both in terms of customer experience as well as in net revenue. Finally, we share the lessons learned from building a comprehensive, modern ranking platform for use in a large-scale e-commerce environment.

\end{abstract}


\ccsdesc[500]{Computing methodologies~Neural networks}
\ccsdesc[500]{Information systems~Recommender systems}

\keywords{Personalization, Recommender Systems, Transformers, Retrieval}

\maketitle

\section{INTRODUCTION}

With the vast choice of items available in e-commerce, finding relevant content has become increasingly challenging. This is why personalization is crucial in showcasing products that align with customers' preferences and session intent. Consequently, large e-commerce companies such as Zalando, one of Europe's largest online fashion e-commerce platforms, are heavily invested in the development of advanced ranking systems that can more effectively cater to customer needs and tastes. 

\begin{figure}[t]
  \centering
  \includegraphics[width=0.7\linewidth]{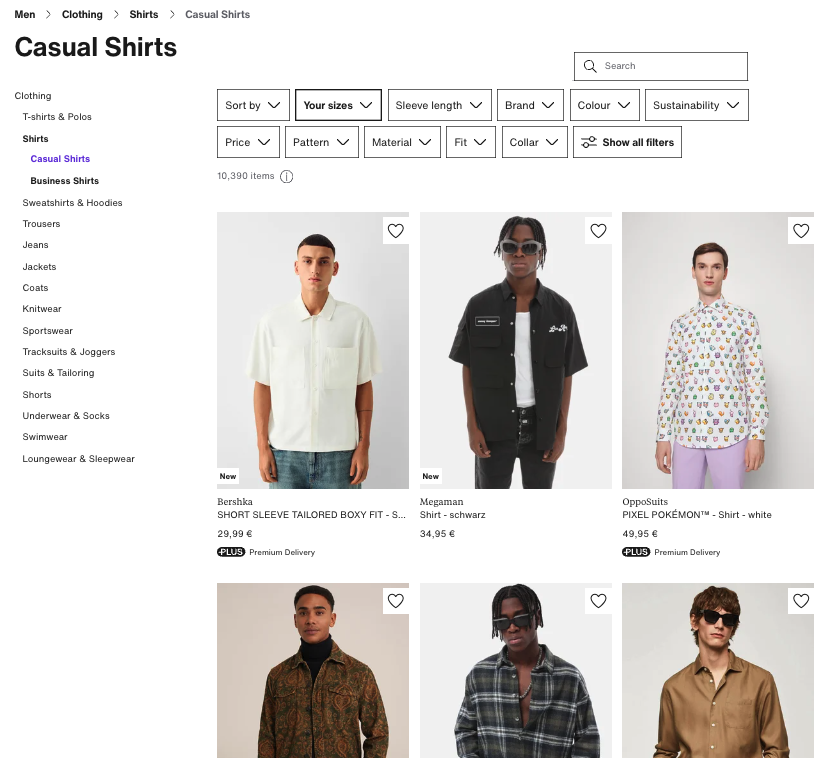}
  \caption{Item catalog browse and search page. On the left is the item category tree, and in the top-right the search query box.}
  \label{fig_browse_search}
\end{figure}

In major e-commerce platforms, deploying larger and more powerful models poses challenges due to the complexities involved in handling high traffic loads in production as these systems must be capable of serving thousands of requests per second across millions of items and customers. Furthermore, browsing and searching the catalog (refer to \autoref{fig_browse_search}) represent primary methods through which customers discover products, whether for immediate purchase or inspiration. However, distinct yet related customer experiences, such as search and browse functions, are often powered by entirely separate systems \cite{Celikik_2022}. This separation increases modeling complexity, increases maintenance costs, and may result in inconsistent customer experiences.


The ability to provide a personalized, real-time, and scalable ranking platform that can be employed across various experiences has become critical to driving customer engagement and business value in e-commerce. We achieve this by building a ranking platform grounded in four key design principles: 1) composability - our platform consists of multiple state-of-the-art ranking models and candidate generators working in orchestration; 2) scalability - ensured by vector-based indexing and scoring; 3) shared real-time serving infrastructure and 4) steerable ranking that can adapt to varying customer preferences and business objectives. 

Our platform is able to support the integration of multiple models through vertical layering, and horizontal integration, blending the outputs of various models or other candidate sources. This enables scalability, independence, and ability to mix various content types to cater to specific use cases as well as building ranking ensembles by combining the outputs of multiple models. The scalability of the first or candidate generation layer is ensured by employing a vector store, facilitating efficient indexing, scoring, and retrieval which are crucial for managing a growing catalog of items and number of customers. To this end, we compute dense representations of customer behaviors, contexts, and item inputs in a common embedding space. Built on a foundation that allows for near real-time scoring and computation of customer and item representations, our platform dynamically adapts the ranking in all layers to recent customer changes which increases the probability of discovering relevant items \cite{transact,qi2020searchbased,bst}. Moreover, utilizing efficient transformer-based model architectures across all layers allows sharing of the serving infrastructure, which in turn reduces engineering complexity, increases re-usability, and helps avoid inconsistency between training and serving phases, which is a common issue in machine learning engineering systems \cite{trainserveskew}. 





Many works both from industry and academia do not try to capture the entire customer journey and side information such as item metadata, customer profile, and contextual inputs \cite{bst,sasrec,bert4rec}. However, it is known that deep learning recommender systems live up to their full potential only when numerous features of heterogeneous types are included \cite{netflixdl}. We demonstrate that incorporating heterogeneous inputs that capture the full spectrum of the customer journey (customer behavior, content-based data, local and global contextual and temporal information) is crucial for ranking quality and even diversity. All of this helps provide contextually relevant results for both in-session browsing, where the customer is actively engaged in shopping, and cross-session scenarios, where the customer returns to the platform after a break with a potentially new shopping intent. To ensure a more streamlined and effective data integration process into the ranking models, contrary to common approaches \cite{bst,transact} that employ additional architectures, our approach utilizes the same self-attention mechanism to efficiently fuse all input data types. 


Many ranking systems in the literature rely on pre-trained items and customer embeddings. Our experiments reveal that similarly to NLP tasks \cite{bert}, the effectiveness of our models significantly increases when the pre-trained input item embeddings are further fine-tuned on the ranking task. Notably, we show that if these embeddings are not continuously trained, the candidate generation model shows substantially less customer engagement. To address the item cold-start problem, we introduce epsilon-greedy exploration by blending fresh items from additional candidate sources into the organic ranking. To address the customer cold-start problem, we leverage customer context and in-session data.


The key contributions of our work are as follows: 

\begin{enumerate}[nosep]

\item We present a comprehensive, flexible, scalable ranking platform able to provide near real-time inference in all ranking layers in high-load systems, building on state-of-the-art models and standard design patterns that can be applied in various search and ranking use cases;

\item We propose novel modifications of existing state-of-the-art ranking model architectures allowing more efficiency without loss of quality. With this we demonstrate that sequence-based models can successfully replace traditional ranking systems in all ranking phases and significantly improve performance; 

\item We present extensive experimentation, including both online and offline. We demonstrate that our proposed system not only significantly outperforms existing solutions by a wide margin (10-40\% improvement in offline evaluation metrics, 15\% combined engagement uplift, and +2.2\% combined net revenue in 4 online A/B tests) but that it also scales effectively under heavy load. 

\end{enumerate}

It's crucial to note that although the experimental results presented are specific to the e-commerce sector, the methodologies, algorithms, and infrastructure discussed are designed for adaptability and can be extended to domains beyond e-commerce. The system has been deployed and operational for the last 12 months in one of the largest e-commerce platforms in Europe. It has successfully replaced numerous legacy systems and it is serving millions of customers per day and handling thousands of RPS.

The remainder of this paper is organized as follows: related work is reviewed in \autoref{section_related_work}. Details on the overall system architecture and design decisions are elaborated in \autoref{section_system_design}. Sections \ref{section_candidate_generation}, \ref{section_ranking_layer} and \ref{section_policy_layer} describe the candidate generation, ranking and policy layers. The experimental results (both offline and online) are presented in \autoref{section_experiments}. Finally, we present the conclusions in \autoref{section_conclusion}.

\section{RELATED WORK}
\label{section_related_work}

Thanks to their advantages over traditional deep-learning-based models, sequence-based recommender systems \cite{sasrec,bert4rec} in their two flavors of language modeling (CLM and MLM) \cite{trm4rec,gsasrec} have gained wide traction. These systems have proven powerful in modeling customer behavior as a sequence of actions due to their capability to 1) capture both short-term and long-term interests \cite{transact}; and 2) the ability to compute complex feature interactions. However, most existing works are trained and evaluated on public datasets that sometimes are not adequate for sequential recommendation tasks \cite{flawsinoffline,sasrec,bert4rec,trm4rec}. Moreovewer, most of the works focus only on offline experiments, with only a few works reporting actual customer impact through end-to-end A/B testing in large-scale environments \cite{transact,pinnerformer}. Our paper extends this line of work and demonstrates the usefulness of these models in real-world applications that include personalized item browsing and search. 
Unlike related approaches \cite{bst,transact}, our approach to scoring candidate items does not suffer from performance issues caused by a long sequence length due to concatenating the candidate item embeddings with the customer action sequence embeddings as in \cite{bst} or limiting the input embedding size as in \cite{transact}. We do not observe a significant drop in ranking performance when feeding only the average embedding of the candidates as a fixed input position into the transformer network compared to including all candidate items as part of the input sequence.

Moreover, only a few published studies \cite{netflixdl,Celikik_2022,bst,transact,celikik2022reusable} include heterogeneous inputs such as context and content-based features, which help address the cold-start problem and address data sparsity by improving generalization. For example, \cite{bst} considers customer profiles and contextual features by using wide \& deep learning (WDL), which relies on a concatenation of signals in the output of the network, making it inadequate to capture powerful feature interactions. \cite{transact} employs deep and cross-network (DCN) on top of a transformer network to explicitly model feature-crosses, which significantly increases the number of parameters of the model. 

\section{SYSTEM DESIGN}
\label{section_system_design}

\begin{figure*}[t]
  \centering
  \includegraphics[width=0.5\linewidth]{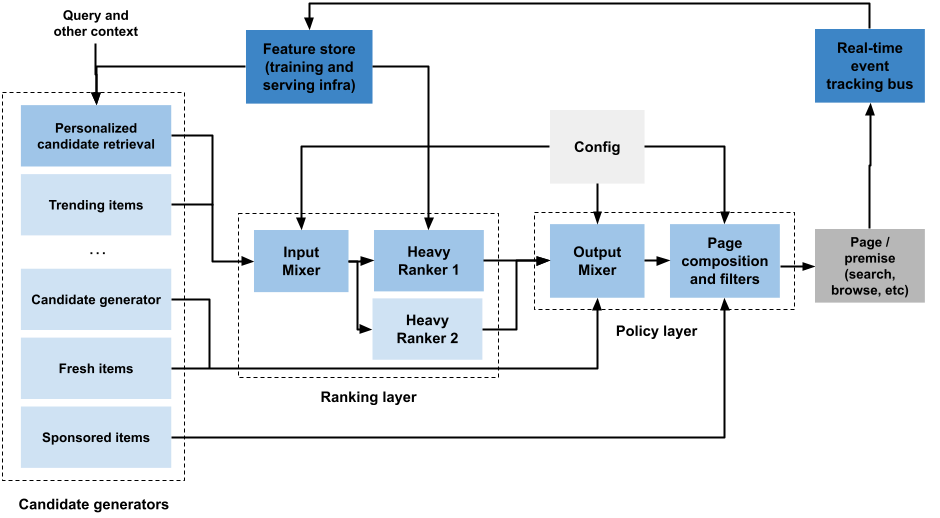}
  \caption{Ranking Platform: Overview}
  \label{fig_ranking_platform}
\end{figure*}

Designing a highly performant, scalable, and steerable ranking platform entails challenges and complex choices. The existing literature often focuses only on subsets of them \cite{mldesignpatterns,youtuberec,twotowersampling,nikerecsys}, while this section aims to navigate through them holistically. We describe foundational design principles and provide an overview of our system architecture and components (see \autoref{fig_ranking_platform}). The presented design is generalizable and applicable to other retrieval and ranking use cases and setups.

\textbf{Composability and orchestration of multiple models.} Our platform enables the combination of various models for several applications. This can be done "vertically", by layering them on top of one another for multi-layered ranking and retrieval which allows scalability and avoids separation of concerns. Ranking models can also be combined "horizontally", by blending the outputs of multiple models or candidate generators. Our platform consists of three layers. The first layer retrieves relevant candidates from multiple candidate generators, possibly generating different content types (e.g. items, outfits, stories, etc.) for use cases such as feeds. Each candidate generator typically entails a (lightweight) ranking model. Subsequent layers refine these selections. The ranking layer applies heavy personalized models for ranking pre-selected items of possibly different types. The policy layer ensures compliance with business or product specifications. Mixing strategies mix outputs to suit specific needs, such as combining different content types in desired proportions or balancing popular, fresh, and personalized content \cite{netflixrecsys}. Model blending is also the blueprint for building ranking ensembles, where outputs of multiple models are combined either by score weighting or meta rankers.


\textbf{Scalable platform.} As already mentioned, a multi-layered model architecture allows high scalability. The more accurate, but computationally heavier part of the ranking is performed on later layers only on a small subset of candidates obtained from the less accurate but more computationally efficient candidate generator layers. Thus, the highest scalability requirements are placed on the candidate generator. Scalability of the candidate generator layer is achieved through a vector store allowing efficient indexing, scoring, and retrieval capabilities to scale for a large and growing item catalog and customer base. Our platform includes infrastructure to compute customer embeddings in near real-time each time a customer accesses the platform. Item embeddings are computed in a streaming-based fashion whenever a new item is introduced and afterward asynchronously indexed. 

\textbf{Shared near real-time serving infrastructure}. Thanks to the similarity of model types and architectures employed in the candidate generation and ranking layers, the training dataset and a large part of our serving infrastructure are reused across the layers. In addition, the use of deep sequence models allows for lightweight feature engineering pipelines consisting of embedding mappings. These mappings reside in the model graph, which guarantees efficiency and consistency between training and serving. Our online feature store is shared between the different ranking layers allowing effective caching of inputs for a given ranking request as well as decreasing the engineering complexity.

\textbf{Steerable ranking}. Our system's flexibility allows for external adjustments to ranking objectives via multi-objective optimization to align with business goals such as customer engagement and revenue. It supports ranking and mixing diverse content types through its candidate generators and mixing components. Finally, business heuristics are applied in the policy layer.

\section{CANDIDATE GENERATION LAYER}
\label{section_candidate_generation}

The objective of our candidate generation layer is to generate personalized item candidates from the item catalog for each individual customer efficiently in near real-time by scoring a vocabulary of millions of items. According to our findings, a personalized and context-aware candidate generator is essential for the performance of the overall ranking system. The top-500 candidates are then reranked by a heavy ranking model in the ranking layer. 

\begin{figure*}[t]
  \centering
  \includegraphics[width=0.5\linewidth]{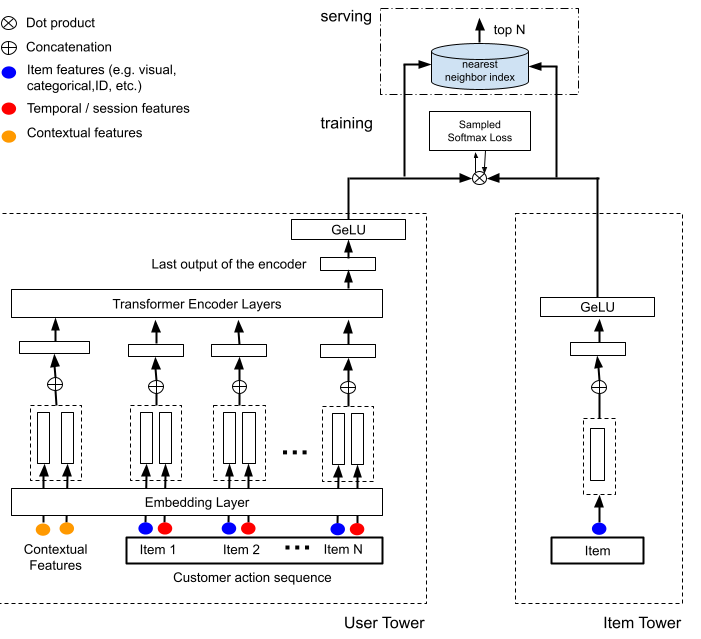}
  \caption{A two-tower model in candidate generation layer used to learn (customer, context) and item embeddings.}
  \label{fig_two_tower}
\end{figure*}

We follow the classical two-tower approach \cite{twotowersampling2}, where the customer tower processes historical customer action sequences and contextual data to generate a customer embedding while the item tower is responsible for generating item embeddings. These embeddings are then combined by using dot product to generate a score per item as shown in \autoref{fig_two_tower} (a theoretical justification about the expressiveness of the two-tower model is provided in the Appendix, \autoref{section_theorem_two_tower}).

We formulate the retrieval task as an extreme multi-class classification problem \cite{youtuberec} with softmax optimization. Every item in the vocabulary represents a distinct class, and the goal is to accurately predict the class of the next item a customer will interact with. We employ sampled softmax loss with log-uniform sampling, with negative classes that correspond to 0.42\% of the total number of classes. This loss outperformed other loss functions and negative sampling strategies. Specifically, we experimented with “generalized” binary cross entropy \cite{gsasrec} and popularity sampling with varying numbers of negatives as well as sampling hard negatives from the category of items the customer was browsing before acting. 

To compute a customer embedding from the customer action sequence and the context, we employ a transformer encoder and use causal language modeling (CLM) as in \cite{sasrec}, processing each customer sequence once per epoch. We predict the subsequent item in the sequence while preventing backward attention by using a causal mask. Scores are computed by multiplying the output vector of the encoder with an output item embedding matrix.

Although trained together, the item and customer towers are deployed and operated independently. The item tower generates item embeddings that are indexed in a vector store. The streaming-based indexing takes place whenever a new item joins the platform, the model is retrained or the inputs of an existing item have been changed. The customer tower is invoked to generate customer embeddings each time a customer accesses the platform. This follows a call to a feature store to fetch all customer-specific inputs, such as context and behavioral sequence. These are then stored in a cache for other transformer-based models in the funnel to reuse. The freshly computed customer embeddings are then used to find items with similar embeddings in the nearest-neighbors index.

For simplicity, items in the item tower in \autoref{fig_two_tower} are represented by using a single embedding that jointly encodes product metadata (brand, category, material, etc.) and visual cues. It is worth noting that in this common setting, trainable input embeddings on the ranking task in the candidate generation layer performed substantially better in both offline and online tests. Details on how different types of input signals from the customer journey data (e.g. contextual and customer action sequence data) are encoded in the customer tower of the model are captured in \autoref{ssec:customer_journey}.

\section{RANKING LAYER}
\label{section_ranking_layer}

The objective of the ranking layer is to rank items returned by the candidate generation phase by their relevance to the customer and their context by using a powerful ranking model. We model this task as a pointwise multi-task prediction problem, where we predict the probability of the customer performing any of the following positive actions on a candidate item: click, add-to-wishlist, add-to-cart, purchase given the context and their past behavioral data. If a candidate item is associated with any of these positive actions, we consider it a positive item, otherwise, it is considered a negative. 

\subsection{Model Architecture}

\begin{figure*}[t]
  \centering
  \includegraphics[width=0.5\linewidth]{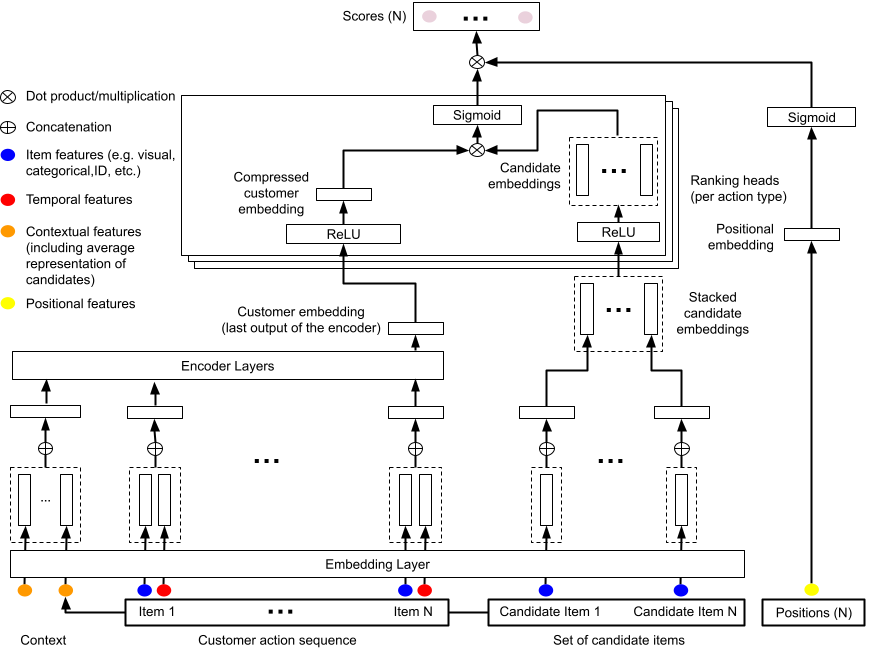}
  \caption{The ranking layer model architecture. The model consists of an embedding layer, transformer encoders, ranking heads (per positive action), a candidate branch, and a position branch used for position de-biasing.}
  \label{fig_ranking_layer}
\end{figure*}

\autoref{fig_ranking_layer} depicts the architecture of the model in the ranking layer. It consists of four main parts: embedding layer, item candidate embedding, customer-context embedding computed via a self-attention mechanism, a prediction head for each of the target action types, and a shallow position branch used for position debiasing. 

For each of the target actions, we define a prediction head, which takes customer and candidate item representations as inputs. Similar as in the candidate generation model, the score per target action is obtained by computing a dot product between the customer-context embedding and all candidate item embeddings in parallel, after passing them through a FFN. A sigmoid function is used to normalize the score and interpret it as a probability. During training, each prediction head contributes equally to the loss, while at serving we produce the final ranking by weighting the scores of each prediction head. The weights are dynamically configurable and determined analytically depending on the customer touch point. 

While models based on list-wise loss directly optimize the ranking objective, the downside is that the predictions from such models do not inherently correspond to probabilities. This lack of calibrated probabilities complicates the multi-objective optimization required for business steering. To this end, we adopt a pointwise loss in our multi-task learning setup. Our underlying assumption is that the tasks share a common internal representation, thereby improving generalization performance through the transfer of knowledge. We employ a cross-entropy loss function utilizing binary relevance labels defined as follows:
\[
    \mathcal{L} = - \frac{1}{N} \sum_{n=1}^{N} \sum_{h=1}^{H} \Big(
        y_n^h \log\left(f^h(x_n)\right) + (1 - y_n^h)\left(\log\left(1 - f^h(x_n)\right)\right)
    \Big),
\]
where $N$ is the number of training examples, $H$ is the number of heads (tasks), $x_n$ input for training example $n$, $y_n^h$ is the target label $\{0, 1\}$ for training example $x_n$ for task $h$ and $f^h(x_n)$ is the output probability of head $h$.


The customer representation vector is generated by encoding and passing the context and the customer sequence actions through a standard transformer encoder with a look-ahead mask. For efficiency, we use only the output of the last position as the customer and context embedding and pass it to the prediction head. The positional encoding is omitted as it has not been proven effective in this as well as in other works \cite{transact}.

Unlike \cite{transact, bst}, we opt out from concatenating the candidate item embeddings as separate positions in the encoder’s input since we have found this to be a limiting factor of the model's scalability, both during training and serving. This is because the number of candidate items can be typically in the order of many hundreds to thousands for a single request. We instead include an average embedding from all candidate embeddings as a single position in the encoder. Both approaches performed similarly well in our setting in terms of ranking quality.



The training objective of our main candidate generator is based on predicting the next customer action, however, our data training pipelines are configurable to predict actions in a longer future time window by masking actions in immediate distance to balance long and short-term customer preferences and improve diversity. A simple masking heuristic we employed from \cite{transact} called "random time window mask" proved effective and significantly increased the number of explored item categories in an A/B test.

\subsection{Encoding of the Customer Journey Data}
\label{ssec:customer_journey}

In this section, we elaborate on how we encode holistic customer journey data into models in both layers. This consists of (1) \emph{behavioral data} which includes customer action sequences, item and action metadata, and temporal data, and \emph{global and local contextual data}. We argue and demonstrate in our experiments that providing complete and heterogeneous information to the model is important for predictions that are personalized and contextually relevant.

\emph{Behavioural data.} As already mentioned, customer action sequences are encoded using transformer encoders as these have been proven more effective than other approaches \cite{transact}. Each action in an action sequence is represented by item embedding that encodes domain-specific visual information, categorical item metadata, action type, and timestamp embeddings. The categorical item metadata consists of relevant attributes such as brand, color, pattern, category, material, etc. The timestamp embedding encodes quantized timestamps as measured from the beginning of the model training. Temporal data is crucial for modeling customer behavior across sessions. Since customer intent can vary drastically across different sessions, modeling action sequences while ignoring this structure affects performance negatively \cite{timeinsecrecsys,celikik2022reusable}. All inputs are passed through trainable embedding to project discrete or bucketized values into low-dimensional spaces. Item and action-specific embeddings are concatenated with the item embedding.  

\emph{Contextual data.} The contextual information is divided into global and local contexts. Global context includes information such as the customer's country and device type while local context includes information about the touch point the customer has triggered an action from, for example, item category, search query, carousel type, and even the products shown on the page. To represent multiple items, we average their embeddings to produce a single “summary” embedding that is fed into its own position in the encoder. To fuse contextual and customer action sequence data we employ the same attention mechanism by allocating the starting positions for contextual features. This approach does not require additional networks such as deep and wide or deep and cross networks that can substantially increase the parameters of the model. Instead, it makes use of the self-attention mechanism to compute complex interactions between the inputs as every other position in the sequence can attend to the context independently. Local contextual information is concatenated with the representation corresponding to the previous action. We note that concatenation in many cases can be replaced by averaging to avoid large input dimensionality. 


\subsection{Position Debiasing}

In the context of ranking systems, feedback loops occur when a model influences customer interactions which can lead to biased relevance data. Typically, items ranked higher by the model receive more customer attention, causing position bias. Position bias causes a skewed representation of actual customer preferences which in turn may amplify and degrade model performance over time due to the feedback loop. To address this, we incorporate position information as a feature into a model to separate the effect of the position and the true relevance of the probability with which the customer would interact with an item. A debiased model is conditioned on positions during training (right branch in \autoref{section_ranking_layer}) and position-independent during serving by setting positional feature to fixed values to counter position bias \cite{multitaskranking}. The position branch is separate from the rest of the model due to the asymmetry between training and serving. We performed additional A/B test which confirmed that adding position debiasing resulted in increasing long-tail utilisation by 5.7\% (measured by item popularity at top-6) as well as improved catalog item utilization by 3.1\% (measured by effective catalog size at top-10) while not deteriorating engagement and financial metrics.

\section{POLICY LAYER}
\label{section_policy_layer}

The last stage of the system, the policy layer, is responsible for the final page composition. Here, multiple re-ranked candidate items are combined into one, performing granular, page-level optimization, and applying heuristics, business rules, and filters depending on the use case. In the following, we describe how we promote fresh (cold-start) items while simultaneously introducing exploration into the system. We also describe some common heuristics applied in this layer to meet product requirements. 

\textbf{Exploration with New Items.} To tackle the cold start problem, the policy layer incorporates fresh items into the organic ranking using exploration heuristics, beginning by sorting these items using content-based features. The blending of outputs from different candidate sources is managed through epsilon-greedy exploration, providing flexibility and ensuring a clear separation of tasks. This method adapts to various use cases by allowing for different exploration techniques and criteria for defining fresh items.

Epsilon-greedy exploration, a staple in reinforcement learning, uses a constant exploration factor that, despite some inefficiencies, functions well in practice and scales to complex scenarios \cite{epsgreedy}. Starting from position k, the policy layer introduces new items with a probability of $\epsilon$ and selects from the ranked list with a probability of $1-\epsilon$, based on a weighted random sampling method determined by the ranking layer. The parameters $k$ and $\epsilon$ help balance the introduction of new items against potential disruptions to the user experience (refer to Algorithm \ref{alg_epsilon_greedy} in the Appendix for more details).

\textbf{Business Heuristics}. This section introduces straightforward heuristics addressing i) \emph{down-sorting previously purchased items} and ii) \emph{avoiding perceived lack of diversity}. Items with diminishing returns, such as a winter coat purchased again soon after the initial buy, are down-ranked to enhance the customer experience. Instead of modeling the probability of repurchase for these items \cite{utilityBasedReco13}, we apply a simpler rule: any item purchased within the last 2 months is down-ranked. Additionally, to prevent the impression of uniformity when many items from the same brand are shown together, we use a diversification heuristic: if a sequence \(a_n,\ldots, a_{n+k}\) of \(k\) same-brand items appears, the first differing-brand item in the subsequent sequence \(a_{n+k+1}, \ldots, a_M\) is relocated to position \(n+k\).


\section{Model Productionization}

Although the two-tower model is trained as a single entity, we deployed each tower as separate endpoints. The item tower is triggered when a new model is trained, a new item is added, or an attribute is modified. The newly generated embedding is then transmitted through a Kafka-based intake stream and indexed in ElasticSearch.

One of the most challenging aspects of our infrastructure was updating the embeddings after training a new two-tower model, as it required maintaining consistency between the embedding versions and the tower model versions. When a new model is released, a complete refeed is necessary to index the updated product embeddings. During this refeed, the system continues to operate using the previous versions of the models and product embeddings. The transition between versions is managed through a blue-green deployment strategy.

For scoring items in the candidate generation phase, we utilize ElasticSearch's vector search function, which employs an efficient approximate k-NN search. An action ingestion is employed by using a near real-time feature store with a caching layer used to temporarily store the customer data ready to be fed to ranking models from different layers thanks to their similar model architecture. The product metadata maps are stored within the model; therefore, only IDs are transferred over the network. This improves the throughput and maintains low latency during inference. 

Model serving is performed on standard CPU-based instances hosted on AWS SageMaker, while training is conducted on multiple GPU instances. The average model response p99 latency for each layer is maintained at 10ms. 

\section{EXPERIMENTS}
\label{section_experiments}

In this section, we present the results of extensive online and offline experiments and ablation studies using internal datasets. We compare the newly introduced ranking system against the existing baseline models for browse, search and item recommendation use cases.

\subsection{Offline Experiments}

\subsubsection{Dataset}

The offline dataset consists of a sample of historical item interactions aggregated by customer id. The item interactions consist of product clicks, add-to-wishlist, add-to-cart, and checkout events attributed to the browse and search premise by using the "last-touch" attribution model. These actions are used as implicit feedback for training and evaluation. The interactions are joined with the corresponding item IDs, their timestamps, and interaction type and sorted by timestamp. To form a single data sample, the resulting sequences are combined with contextual data, specifically, market, device type, browsing category, and search query (if present). The training dataset consisted of 71M unique customers across 25 markets. We do not perform any preprocessing such as de-duplication or outlier removal on the obtained customer sequences besides truncating to the last 100 actions. The average sequence length is 24 actions. The evaluation dataset contains 300K customers (\autoref{fig:interactions_histo} in the Appendix provides histograms per action type). We applied hard temporal split to create the training and test datasets to ensure no data leakage.

\subsubsection{Metrics and evaluation protocol}
\label{section_metrics}

The ranking models have been evaluated on following metrics:
\begin{itemize}[nosep]
\item\textbf{Recall@k}: defined as the proportion of all relevant items within top k items (as defined in \cite{offlinemetrics});
\item\textbf{NDCG@k}: measures the effectiveness of a ranking by taking into consideration the position and the relevance label of each item in the ranked list of top-k items (as defined in \cite{offlinemetrics});
\item\textbf{Diversity}: we use the maximum run of consecutive items from the same brand as a proxy to brand diversity. A high value suggests that a ranking can be dominated by items coming from only a few brands. User acceptance testing showed this leads to undesirable customer experience;q
\item\textbf{Novelty}: we use recall of new items as a proxy for novelty. This metric captures the ability of a ranker to promote new items and address the item cold-start problem. 
\end{itemize}



Our evaluation protocol closely mirrors a real-world production environment by employing a strict separation of training and test data based on time. Customer sequences within the dataset are chronologically ordered. The models undergo evaluation exclusively on the test dataset (with time-based split), which comprises “ground-truth” pages of items that users have either viewed or interacted with following search or browse requests. Importantly, models are provided data only up until the timestamp of each request, with a particular emphasis on adhering to data caching periods—during these times, models operate solely on cached data. 

For each item category or search query contained in the requests, the candidate generation models score and ranks all corresponding items in the catalog. The ranking models re-rank the 500 items with highest score coming from the candidate generation layer. We then calculate the offline metrics based on these ranked lists against the above "ground-truth" pages observed in the test data.

The metrics are calculated for each ranking produced for a single test example, and averaged over the test dataset. All reported results are statistically significant (p-value < 0.05) unless stated otherwise. We used a t-test for significance testing. 

\subsubsection{Candidate Generation}

We compare the following methods in the candidate generation layer:
\begin{itemize}[nosep]
\item\textbf{GBT} is a candidate generation model based on Gradient Boosting Trees, which has proven to offer competitive performance compared to neural-based models \cite{gbdt}. It ranks items based on their static metadata (season, material, category type, etc.) and their dynamic historical engagement rates (add-to-cart, add-to-wishlist and click rates for the last 5 min). The LambdaRank objective~\cite{lambdarank06} is used during training to up-rank interactions based on their graded relevance ("purchase" has the highest while "click" the lowest relevance). The model's scores are computed in streaming fashion making it highly reactive to trends in customer behavior. This model was our previous production model.
\item\textbf{RCG} is our candidate generation model introduced in \autoref{section_candidate_generation}. We test a few variants of this model: RCG$_{\mathrm{ntr}}$ which uses pre-trained visual embeddings of items \cite{fdna2016} and includes a bias term that captures item popularity; RCG$_{\mathrm{tr}}$ which employs trainable item embeddings, initialized from the pre-trained visual item embeddings. These two models utilize only global contextual information, such as country and device type. Additionally, RCG$_{\mathrm{tr+ctx}}$ denotes a variant with additional local contextual input --- the current user's browsing category and a binary flag whether they browse or search. All the model versions consist of 2 encoder layers with 4 heads, gelu activation, and a max. sequence length of 100. The model was trained for 20 epochs by using the Adam optimizer, with a learning rate set to 0.001. We do not consider hard negatives and rather employ log uniform candidate samplin as described in \autoref{section_candidate_generation}.
\end{itemize}


\begin{table}[]
\caption{Ablation study of the ranking candidate generation model (RCG) on browse traffic relative to the baseline candidate generator GBT. Recall@500, which is the main offline metric for candidate generation, is shown per customer segment while NDCG is shown for all customers.}
\label{tab:rcg}
\begin{tabular}{l|c|c|c|c}
\toprule
\multirow{2}{*}{\textbf{Model}} & \multicolumn{3}{c}{\textbf{Recall@500}} & \multirow{2}{*}{\textbf{NDCG@500}} \\
 & \textbf{All} & \textbf{New} & \textbf{Returning} & \\
\midrule
RCG$_{\mathrm{ntr}}$ & +16.2\% & +6.00\% & +32.56\% & +31.5\% \\
RCG$_{\mathrm{tr}}$ & +30.1\% & +9.38\% & +67.36\% & +54.2\% \\
RCG$_{\mathrm{tr+ctx}}$ & +43.6\% & +24.4\% & +76.0\% & +76.6\%\\
\bottomrule
\end{tabular}
\end{table}

\begin{table}
    \centering
    \caption{Performance summary of the ranking candidate generation (RCG) models on the search use case relative to the baseline candidate generation GBT model.}
    \label{tab:ls_rcg_cat_search}
    \begin{tabular}{lll}
    \toprule
    \textbf{Model}                   & \textbf{Recall@500} & \textbf{NDCG@500} \\
    \midrule
    RCG$_{\mathrm{tr}}$     & +2.7\%        & +16.8\%     \\
    RCG$_{\mathrm{tr+ctx}}$ & +4.6\%        & +22.0\%     \\
    \bottomrule
    \end{tabular}
\end{table}

\autoref{tab:rcg} presents an offline ablation study comparing our RCG model with the baseline GBT used as a candidate generator on the browse use case while \autoref{tab:ls_rcg_cat_search} summarizes the offline evaluation results of the RCG model on the search use case only. Similar improvements seen on the browse traffic apply to the search traffic. In summary, across all evaluated customer segments, our newly proposed candidate generator significantly outperforms the existing one.

The model variant incorporating trainable item embeddings achieves markedly improved performance in offline metrics. However, it is important to note that trainable item embeddings exacerbate the item cold-start problem (to mitigate this issue, we introduce exploration strategies for new items in \autoref{section_policy_layer}).

Incorporating local contextual inputs further boosts the performance of the RCG$_{\mathrm{tr}}$ model. Specifically, by including item category and search query presence, the RCG$_{\mathrm{tr+ctx}}$ model's Recall@500 is enhanced by an additional $10\%$, and its NDCG by $14\%$. In \autoref{section_online_experiments} we will show that these gains are also reflected in our online experiments.


\subsubsection{Ranking Layer}

\begin{table}[]
\caption{Relative performance change compared to WDL-ATT, on relevance, freshness, diversity and novelty.}
\label{tab:rl_offline}
\begin{tabular}{c|l|c|c|c}
\toprule
\textbf{k} & \textbf{Model} & \textbf{NDCG} & \textbf{Novelty} & \textbf{Diversity} \\
\midrule
\multirow{2}{*}{6} & RL & +11.76\% & +10.15\% & +1.92\% \\
 & BST & -11.05\% & +7.50\% & +8.81\% \\
\midrule
\multirow{2}{*}{84} & RL & +5.94\% & +1.16\% & +6.18\% \\
 & BST & -3.90\% & +1.41\% & +10.99\% \\
\bottomrule
\end{tabular}
\end{table}

We compare the following methods in the ranking layer:
\begin{itemize}[nosep]
\item\textbf{WDL-ATT} is a wide \& deep neural network with an attention mechanism between the customer action sequences and ranking candidates. It scores the candidate items by applying a dot product between context, user, and item embeddings that are all trainable and have 128 dimensions. It employs a loss function directly optimizing for the NDCG metric~\cite{approxnDCG19}. Items are represented by using pre-trained embeddings that encode brand, category, pattern, and other visual cues. It is trained for 2 epochs by using the Adam optimizer with a learning rate of $0.0002$.
\item\textbf{RL} is our ranking model introduced in \autoref{section_ranking_layer}. The model consists of 2 encoder layers with 8 heads, relu activation, and a maximum sequence length of 80. $d_{\textrm{model}}$ and the model output size is set to 128. The model is trained for 2 epochs by using the Adam optimizer, with a learning rate of $0.001$. The ratio of negative vs. positive samples is set to 4. Since sampling negatives only from items that were in the view-port lead to degradation of performance, we sampled from all non-interacted items on the page. 

\item\textbf{BST} is the Behavior Sequence Transformer introduced in \cite{bst}. We use the same inputs and hyperparameters as in RL.
\end{itemize}

It should be noted that all compared algorithms use the same near real-time serving infrastructure. Some of the algorithms mentioned in related work, such as TransAct \cite{transact}, while potentially competitive, were not applicable to our use case due to latency constraints. In these experiments, we focus on the NDCG metric for "high-value actions" or HVAs, which, in our context, are add-to-wishlist and add-to-cart actions. This metric acts as a proxy for our success KPI, defined by customer engagement wrt. HVAs which is described in more detail in \autoref{section_online_experiments}.

The offline evaluation results are summarized in \autoref{tab:rl_offline}, comparing RL and BST against the existing WDL-ATT which proved to be a strong baseline. The NDCG metric indicates that the new model effectively prioritizes relevant items higher up in the rankings, both at the top of the list ($k=6$) and across the entire first page ($k=84$). Furthermore, RL favored the promotion of new items while enhancing diversity. In terms of relevance, BST lagged behind both algorithms, although it performed the best when it comes to diversity. 

\autoref{tab:rl_ablation} shows an ablation study that describes the contribution of each input type to the model's performance. The removal of any of the inputs significantly affects the overall model accuracy. Particularly noticeable is the performance decline when contextual inputs are not integrated early in the encoder. This finding suggests that our model leverages contextual information more effectively alongside rich item representations in customer action sequences compared to algorithms such as BST that fail to capture complex interactions between contextual, action and item metadata. 
Additionally, the ablation study shows that omitting heterogeneous inputs substantially diminishes the model's performance. Specifically, excluding contextual inputs results in more than a 10\% decrease in NDCG@6, while complete removal of item metadata leads to a drastic 26\% reduction in NDCG@6.

\begin{table}[]
\caption{Ablation study: relative change of the NDCG metric at $k=\{6,84\}$, measured after removing various inputs in the RL model.}
\label{tab:rl_ablation}
\centering
\begin{tabular}{l|c|c}
\toprule
\multirow{2}{*}{\textbf{Modification}} & \multicolumn{2}{c}{\textbf{NDCG}} \\
& \textbf{k=6} & k=84 \\
\midrule
w/o context & -9.35\% & -3.96\% \\
w/o candidates in the context & -2.72\% & -1.51\% \\
w/o context in encoder & -13.81\% & -5.45\% \\
w/o customer history & -23.64\% & -11.07\% \\
w/o visual embeddings & -8.94\% & -3.82\% \\
w/o categorical item metadata & -3.09\% & -1.43\% \\
\begin{tabular}[c]{@{}l@{}}w/o visual embeddings \\ and categorical item features\end{tabular} & -16.68\% & -7.28\% \\
w/o item id embeddings & -13.69\% & -5.61\% \\
\bottomrule
\end{tabular}
\end{table}

\subsection{Online Experiments}
\label{section_online_experiments}

\begin{table*}[]
\caption{Summary of A/B tests performed on the browse use case. The relative percent uplifts are reported with engagement being the success KPI. The best-performing variant in each test is highlighted in bold. Non-significant results are marked with *.}
\label{tab:rcg-ab}
\centering
\begin{tabular}{l|l|c|c|c|c}
\toprule
 & \textbf{Variant} & \textbf{Engagement} & \textbf{Revenue} & \textbf{Novelty} & \textbf{Diversity} \\
\midrule
\multirow{3}{*}{1} & GBT + WDL-ATT & - & - & - & - \\
& \textbf{RCG$_{\mathrm{tr}}$} & \textbf{+4.48\%} & \textbf{+0.18\%*} & \textbf{-19.5\%} & \textbf{-31.2\%} \\
& \begin{tabular}[c]{@{}l@{}}RCG$_{\mathrm{ntr}}$ + WDL-ATT\end{tabular} & +1.61\% & -0.46\% & -18.5\% & -8.5\% \\
\hline
\multirow{3}{*}{2} & RCG$_{\mathrm{tr}}$ & - & - & - & - \\
& RCG$_{\mathrm{tr}}$ + WDL-ATT & +1.51\% & +0.21\% & +14.8\% & +14.3\% \\
& \textbf{RCG$_{\mathrm{tr}}$ + RL + PL} & \textbf{+4.04\%} & \textbf{+0.86\%} & \textbf{+46.3\%} & \textbf{+33.7\%}\\
\hline
\multirow{2}{*}{3} & RCG$_{\mathrm{tr}}$ + RL + PL & - & - & - & - \\
& \textbf{RCG$_{\mathrm{tr+ctx}}$ + RL + PL} & \textbf{+2.40\%} & \textbf{+0.60\%} & \textbf{+12.1\%} & \textbf{+6.4\%}\\
\bottomrule
\end{tabular}
\end{table*}

\begin{table*}[]
\caption{Summary of A/B tests performed on the search use case. The relative percent uplifts are reported with engagement being the success KPI. The best-performing variant in each test is highlighted in bold. Non-significant results are marked with *.}
\label{tab:rcg-ab-search}
\begin{tabular}{l|l|c|c}
\toprule
 & \textbf{Variant} & \textbf{Engagement} & \textbf{Revenue} \\
\midrule
\multirow{2}{*}{1} & GBT$_{\mathrm{tr}}$ + WDL-ATT & - & - \\
& \textbf{RCG$_{\mathrm{tr}}$ + RL + PL} & \textbf{+3.11\%} & \textbf{+0.15\%*}\\
\hline
\multirow{2}{*}{2} & RCG$_{\mathrm{tr}}$ + RL + PL & - & - \\
& \textbf{RCG$_{\mathrm{tr+ctx}}$ + RL + PL} & \textbf{+0.70\%} & \textbf{+0.17\%}\\
\bottomrule
\end{tabular}
\end{table*}

\begin{table*}[]
\caption{Summary of A/B tests performed on recommendation use cases. The relative percent uplifts against the baseline are reported with engagement being the success KPI. The best-performing variant in each test is highlighted in bold.}
\label{tab:rl-reco}
\begin{tabular}{l|l|c}
\toprule
\textbf{Reco use case} & \textbf{Variant} & \textbf{Engagement} \\
\midrule
\multirow{2}{*}{PDP (similar items)} & CF (baseline) & - \\
& \textbf{CF + RL} & \textbf{+0.78\% HVA} \\
\hline
\multirow{2}{*}{Home (product campaign)} & Curated (baseline) & - \\
& \textbf{RL} & \textbf{+21.2\% CTR} \\
\bottomrule
\end{tabular}
\end{table*}

We have conducted several online A/B tests on real-world ranking and recommendation use cases at Zalando. Whenever possible, these tests were carried out systematically, replacing one component at a time to evaluate its impact. All tests allocated equal user splits among variants (a given users always remains in the initially allocated variant) over a few weeks, as necessary, to achieve the minimum detectable effect for the success KPI with a $p$-value$< 0.05$. Each model was retrained and deployed daily. Beyond customer engagement, our evaluation of online performance sometimes includes exploratory metrics, including financial metrics, the capacity to promote new items (novelty), and ranking diversity (as defined in \autoref{section_metrics}). 

\autoref{tab:rcg-ab} summarizes the results from a series of A/B tests performed on the browse use case, where a customer is browsing the category tree as depicted in \autoref{fig_browse_search}. In the first A/B test in \autoref{tab:rcg-ab} (row 1), we compared the new candidate generation model (RCG) against the previous ranking system end-to-end (GBT + WDL-ATT). We evaluated two variants: firstly, the RCG model with trainable item embeddings (RCG${\textrm{tr}}$) and, secondly, the RCG model with non-trainable item embeddings (RCG${\textrm{ntr}}$) paired with WDL-ATT as the next ranker in the funnel. The former variant demonstrated a significant increase in engagement. However, the new candidate generation model had no significant impact on financial KPIs, promoted fewer new items, and decreased diversity.

In the second A/B test in \autoref{tab:rcg-ab} (row 2), we evaluated the effect of adding our new ranking layer (RL) along with the policy layer (PL) as the next ranker. The baseline for this experiment was the winning variant from the first test, namely RCG$_{\textrm{tr}}$. For completeness we included the previous ranking algorithm WDL-ATT as another variant. The outcomes highlight the advantages of applying our powerful ranking model on candidates from the candidate generation layer. This variant significantly improved all monitored KPIs across all customer segments, including net revenue per customer, novelty, diversity and even customer retention.

In the third A/B test in \autoref{tab:rcg-ab} (row 3), we tested an improved version of the candidate generation model aka RCG$_{\textrm{tr+ctx}}$, which includes local contextual data. This resulted in a significant uplift in all monitored KPIs, including brand, and categorical diversity (omitted due to space limitations). This A/B test demonstrates the importance of including data that captures the entire customer journey to provide a more contextually relevant ranking.

\autoref{tab:rcg-ab-search} summarizes the results of a series A/B tests performed on the search use case, i.e., when the customer is using full-text search to find their desired item(s). The conclusions are similar and for brevity we omit the details.

\autoref{tab:rl-reco} summarizes the outcome of two A/B tests where RL was used to re-rank existing recommendation baselines. In the PDP (product detail page) use case, RL was used to re-rank the top-200 most similar items produced by k-NN item-to-item collaborative filtering (CF) out of which the top-15 are shown in the "similar item" carousel. In the Home use case, RL was used to rerank a curated set of products from an active campaign and show these items in this order to the customer. As a conclusion, RL can be re-used to personalize other customer touch points.

\subsection{Serving Latency}

Processing of rich contextual and sequence information in the models can increase their latency, making them unsuitable in real-world applications. While we have seen an increase in the overall latency, the extra latency in our ranking comes from: i) introduction of the candidate generator endpoint, adding 10ms p99 latency on average for customer embeddings, and ii) a k-NN search in the vector database, adding 30ms p99 latency on average. Our new ranking model has similar inference latency to the legacy WDL-ATT model which is 10ms p99 on average. The total added latency was approximately 40ms, resulting in 200ms end-to-end latency, well below our 500ms SLO. All endpoints were deployed on standard CPU instances.

\section{CONCLUSION}
\label{section_conclusion}

In this paper, we have introduced a flexible, scalable, steerable, and real-time ranking platform that has been proven to enhance customer experience by delivering more relevant and personalized content for various use cases including browse, search and item recommendation. The approach has led to improvements in both customer-centric and business metrics. We have described the architecture of our ranking platform, adhering to a set of design principles and utilizing state-of-the-art models. We have also provided insights into their performance, highlighting the considerable advantages of integrating heterogeneous signals and inputs that encompass the entire customer journey, as well as the effectiveness of fine-tuning input embeddings to boost model performance.

Our offline and online evaluations clearly demonstrate that our proposed system not only significantly outperforms existing solutions by a wide margin (10-40\% improvement in offline evaluation metrics and a 15\% combined engagement and +2.2\% revenue uplift in 4 online A/B tests) but also excels in real-life use cases and scales effectively under heavy load, which is a crucial requirement for large e-commerce platforms. Furthermore, we illustrate that the enhanced experience benefits both returning and new customers. Lastly, we provide valuable insights and practical guidance for application by other applied scientists and practitioners within the domain.




\bibliographystyle{ACM-Reference-Format}
\bibliography{ranking_platform}

\newpage

\section{APPENDIX}

\subsection{Exploration with New Items}
To promote new items that may suffer from the cold start problem, we follow the Algorithm~\autoref{alg_epsilon_greedy} to combine organic ranking results ($S$) coming from retrieved candidates by e.g. RCG, with new items ($N$). The algorithm is controlled by $k$ and $\epsilon$.
\begin{algorithm}[H]
    \caption{Mixing organic and new item candidate sources.}
    \label{alg_epsilon_greedy}
    \begin{algorithmic}
        \For{position $p = 1, \dots, k - 1$}
            \State Select the next item $I \gets \argmax_{i \in S} \text{rel}(i)$
            \State $S \gets S \setminus I$
        \EndFor
        \For{position $p = k, \dots, |S| + |N|$}
            \State $X \sim \text{Bernoulli}(\varepsilon)$
            \If{$X = 1$}
                \Comment{Exploration phase\;}
                \State For each $i \in N,\; w_i \gets \text{rel}(i) / \sum_{j \in N} \text{rel}(j)$
                \State Select the next item $I$ by sampling with the weights $(w_i)_{i \in N}$
                \State $N \gets N \setminus I$
            \Else
                \Comment{Exploitation phase}
                \State Select the next item $I \gets \argmax_{i \in S} \text{rel}(i)$
                \State $S \gets S \setminus I$
            \EndIf
        \EndFor
    \end{algorithmic}
\end{algorithm}

\subsection{Data distribution}
\begin{figure}[H]
    \centering
    \includegraphics[width=0.8\linewidth]{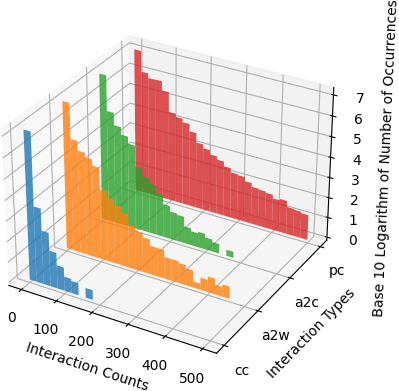}
    \caption{Histogram of the number of occurrences of interaction types used in the user sequences (cc for purchases, a2w for add-to-wishlist, a2c for add-to-cart, and pc for product click). Occurrences are on the logarithm scale.}
    \label{fig:interactions_histo}
\end{figure}

Our datasets are based on actions that customers perform against items on the platform. In \autoref{fig:interactions_histo} we show the distribution of those actions, computed from raw customer data, before any dataset-specific preprocessing, e.g. trimming. Clicks have the biggest volume, then add-to-wishlist, add-to-cart, and purchases.

\subsection{Expressive Power of Two-Tower Models}
\label{section_theorem_two_tower}
The two major recommendation models used in our ranking platform have a two-tower architecture in which one tower embeds the customer and the other one embeds the fashion article that is being scored (see Section~\ref{section_candidate_generation} and Section~\ref{section_ranking_layer}). Mathematically, the score function $f^{model}$ corresponding to a model of this type can be written as
\begin{equation}\label{eq:dotproduct}
    f^{model}(x) = \langle \varphi(c),\,\psi(a) \rangle,
\end{equation}
where $x = (c, a)$ is the model input with $c$ and $a$ being the customer and the article parts of the input, respectively. 

In this section, we study the expressive power of this model class. Specifically, we prove that any continuous target function $f$ (defined on a bounded feature space) can be approximated by a score function of the form \eqref{eq:dotproduct} provided the embedding size is large enough.

\begin{theorem}
    Let the range of customer and article features be bounded: $C^\ell_i \le c_i \le C^u_i$, $i=1,\,\ldots,\,k_c$, and $A^\ell_j \le a_j \le A^u_j$, $j=1,\,\ldots,\,k_a$, and let the target function $f$ be continuous on the feature domain 
    $$
        \mathrm{D} = [C^\ell_1,\,C^u_1]\times\ldots\times [C^\ell_{k_c},\,C^u_{k_c}]\times [A^\ell_1,\,A^u_{1}]\times\ldots\times [A^\ell_{k_a},\,A^u_{k_a}].
    $$

    Then for any $\varepsilon > 0$, there exist $n>0$ and transformations $\varphi\colon \mathrm{R}^{k_c}\mapsto\mathrm{R}^n$ and $\psi\colon \mathrm{R}^{k_a}\mapsto\mathrm{R}^n$ such that

    \begin{equation}\label{eq:approximation}
        \max_{(c,\,a)\in \mathrm{D}} \left|f(c,\,a) - \langle \varphi(c),\,\psi(a) \rangle\right| < \varepsilon.
    \end{equation}
    \begin{proof}
        Without loss of generality, let us assume that $\mathrm{D}$ is a unit cube, i.e. $C^\ell_i = 0$, $C^u_i = 1$ for all $i=1,\,\ldots,\,k_c$ and $A^\ell_j = 0$, $A^u_j = 1$ for all $j=1,\,\ldots,\,k_a$.

        Consider the set of all multivariate polynomial functions on $\mathrm{D}$. Note that it (a) contains constant functions, (b) is closed under the operations of addition and multiplication, and (c) separates points: for any $u,\,v\in\mathrm{D}$, $u\neq v$, there exists a polynomial $P$ such that $P(u)\neq P(v)$. Then by applying the Stone-Weierstrass theorem, we conclude that for any $\varepsilon > 0$, there exists a polynomial $P_\varepsilon$,

        \begin{equation}\label{eq:polynomial}
            P_\varepsilon = \sum_{m=1}^n 
            \alpha_m \prod_{i=1}^{k_c} c^{p_{m,
            ,i}}_i \prod_{j=1}^{k_a} a^{q_{m,
            ,j}}_j,
        \end{equation}
        such that
        \begin{equation}\label{eq:polynomialApproximation}
            \max_{(c,\,a)\in \mathrm{D}} \left|f(c,\,a) - P_\varepsilon(c,\,a)\right| < \varepsilon.
        \end{equation}
        By defining
        
        $$
        \begin{array}{c}
            \varphi(c) = \left(
            \alpha_1 \prod_{i=1}^{k_c} c^{p_{1,\,i}}_i,\,\ldots,\,\alpha_n\prod_{i=1}^{k_c} c^{p_{n,\,i}}_i\right),\\ 
            \psi(a) =  \left(\prod_{j=1}^{k_a} a^{q_{1,\,j}}_j,\,\ldots,\,\prod_{j=1}^{k_a} a^{q_{n,\,j}}_j\right),
        \end{array}
        $$
        we can rewrite \eqref{eq:polynomial} as
        $$
            P_\varepsilon(c,\,a) = \langle \varphi(c),\, \psi(a) \rangle,\quad (a,\,c)\in\mathrm{D}.
        $$
        Then \eqref{eq:polynomialApproximation} implies that the constructed transformations $\varphi$ and $\psi$ satisfy \eqref{eq:approximation}. 
    \end{proof}
\end{theorem}

\end{document}